\documentclass[aps,pra,superscriptaddress,reprint,showkeys,showpacs]{revtex4-1}
\usepackage{amsmath}    % need for subequations
\usepackage{graphicx}   % need for figures
\usepackage{array}
\usepackage[dvipdfx,cmyk]{xcolor}     % use if color is used in text
\usepackage[caption=false]{subfig}  % use for side-by-side figures
\usepackage[colorlinks,linkcolor=red,citecolor=brown,urlcolor=blue]{hyperref}
\usepackage{dsfont,bm}

\newcommand{\be}{\begin{equation}}
\newcommand{\ee}{\end{equation}}
\newcommand{\ben}{\begin{eqnarray}}
\newcommand{\een}{\end{eqnarray}}
\newcommand{\bes}{\begin{subequations}}
\newcommand{\ees}{\end{subequations}}
\newcommand{\bF}{\begin{figure}}
\newcommand{\eF}{\end{figure}}

\begin{document}

\title{Chaotic spin-spin entanglement on a recursive lattice}

\author{Levon Chakhmakhchyan}
\affiliation{A.I. Alikhanyan National Science Laboratory, Alikhanian Br. 2, 0036 Yerevan, Armenia}
\affiliation{Institute for Physical Research, 0203 Ashtarak-2, Armenia}
\affiliation{Laboratoire Interdisciplinaire Carnot de Bourgogne, UMR CNRS 6303, Universit\'{e} Bourgogne Franche-Comt\'{e}, F-21078 Dijon Cedex, France}
\author{St\'{e}phane Gu\'{e}rin}
\affiliation{Laboratoire Interdisciplinaire Carnot de Bourgogne, UMR CNRS 6303, Universit\'{e} Bourgogne Franche-Comt\'{e}, F-21078 Dijon Cedex, France}
\author{Claude Leroy}
\affiliation{Laboratoire Interdisciplinaire Carnot de Bourgogne, UMR CNRS 6303, Universit\'{e} Bourgogne Franche-Comt\'{e}, F-21078 Dijon Cedex, France}

\date{\today}

\begin{abstract}

We propose an exactly solvable multisite interaction spin-1/2 Ising-Heisenberg model on a triangulated Husimi lattice for the rigorous studies of chaotic entanglement. By making use of the generalized star-triangle transformation, we map the initial model onto an effective Ising one on a Husimi lattice, which we solve then exactly by applying the recursive method. Expressing the entanglement of the Heisenberg spins, that we quantify by means of the concurrence, in terms of the magnetic quantities of the system, we demonstrate its bifurcation and chaotic behavior. Furthermore, we show that the underlying chaos may slightly enhance the amount of the entanglement, and present on the phase diagram the transition lines from the uniform to periodic and from the periodic to chaotic regimes.

\end{abstract}

\maketitle

\section{Introduction}\label{intr}

Quantum entanglement, that stands for the intrinsic nonlocal correlations inherent in the quantum theory, plays a crucial role in the understanding of the fundamentals of quantum physics as well as serves as the key resource for implementing practical applications within the field of quantum technologies. Specifically, entanglement is vital to diverse secure quantum communication and teleportation schemes \cite{QKD1, teleport} and is at the heart of quantum computation, quantum metrology, quantum imaging \cite{comp, image, metr}, etc.

Various protocols, applicable to continuous- and discrete-variable systems have been proposed recently, allowing one to generate entanglement shared between quantum memories \cite{memory} and encoded in states of light fields \cite{cv} or, alternatively, to create systems of strongly entangled superconducting or cavity QED qubits, quantum dots, highly excited Rydberg atoms \cite{jj, me1, ry}, etc. Meanwhile, solid-state systems and, particularly, magnetic materials, are of significant importance in this respect as they appear to be a source of entanglement too, even on a macroscopic level \cite{exp, diam}. Furthermore, entanglement turns out to be a powerful tool here for the characterization of quantum phase transitions as it encodes all the information shared between subparts of a system that may fail to be captured by means of ordinary correlation functions (this can be the case in some exotic quantum ground states \cite{hidden}). In this respect, spin models are widely used to describe the properties of such solid-state systems where intercoupled (e.g., by means of the Heisenberg exchange interaction) spins are nested in the sites of a specific lattice structure. The entanglement features of various spin-lattice models of this kind have been thoroughly studied recently \cite{fractal, ananik13, me222, other}, revealing, particularly, a strong connection of magnetic properties and entanglement \cite{witn, witn1, me22} that allows one to witness the latter experimentally (e.g., by means of heat capacity and magnetic susceptibility measurements \cite{exp, exp1}).

Meanwhile, another intriguing question that we address in this paper, is the connection of quantum entanglement and chaos. As is known, the latter also plays an important role in various areas of research, and particularly in the field of quantum information. For instance, the uncontrolled interactions between the qubits of a quantum computer, being above some critical strength, induce chaotic behavior that may break down the efficiency of the computer or bring about exponential decay of the entanglement \cite{dima1, dima2}. Consequently, the complexity of error correction codes increases and the correction time, however, is independent of whether the system is chaotic or not \cite{more}. On the other hand, quantum chaos may also ensue in specific quantum protocols, such as the Grover's search algorithm, or quantum Fourier transform, giving rise to a peculiar combination of quantum signatures of chaos and integrability \cite{braun}. Meanwhile, another suitable platform for the combined studies of entanglement and chaotic behavior is nonlinear dissipative oscillators \cite{kruch} and so-called coupled tops \cite{top1}, where the underlying classical chaos may enhance and determine the amount of the entanglement shared between the coupled systems \cite{ind, top2}. Moreover, quantum signatures of chaotic behavior in a kicked top have been demonstrated for trapped caesium atoms, showing that if prepared in the chaotic sea, the electron and nuclear spins of the atoms get rapidly entangled \cite{exptop}. Finally, some other studies, e.g., for $N$-atom Jaynes-Cummings model have also shown that the entanglement rate is considerably enhanced for chaotic initial conditions \cite{dicke}.

On our part we propose a specific spin-lattice model to study the relation among its chaotic behavior, entanglement, and magnetic properties. Specifically, the structure that we consider is of recursive nature and consists of two non-equivalent types of sites: the nodal ones that constitute a Husimi tree, and trimeric units that form embedded triangles [see Fig.~\ref{lattice}(a)]. Being constructed in such a way, it is thus natural to call this structure a triangulated Husimi tree and its deep interior a triangulated Husimi lattice (THL) (we note that a similar system has been proposed very recently in Ref.~\cite{strecka}). Next, taking into account that the rigorous treatment of purely Heisenberg models are mostly unattainable, we adopt here a spin-1/2 Ising-Heisenberg model on a THL such that the Ising spins are nested on the nodal sites of the lattice, while the Heisenberg variables are situated on the vertices of the embedded triangles. A distinguishable feature of this model is that it can be solved exactly by combining the generalized star-triangle transformation \cite{YD, YD1} and the recursive method \cite{pretti, pretti1, monroe, anan, comm, jetp, lee}. Even more, if three-site interactions between nodal Ising spins are included, the model exhibits chaotic behavior (this is rather different from the case studied in Ref.~\cite{strecka} where only two-site interactions are considered in the absence of the external magnetic field). Although this fact has been well known for an Ising model on a simple Husimi lattice \cite{comm, jetp, monroe}, it brings about drastic changes in the properties of the Heisenberg trimer of the THL. Namely, at sufficiently low temperatures, its magnetization, and even more interestingly, the entanglement (that we quantify by means of the concurrence \cite{wooters}) exhibit period-doubling bifurcations, chaotic behavior, and periodic windows. It is worth noting, however, that the ground state of the Heisenberg triangle remains rigid and that it is the Ising sublattice (to which the triangles are coupled) that gives rise to chaos. Furthermore, as is shown below, the spin-spin entanglement of the triangle may be slightly enhanced by means of the underlying chaotic behavior when compared to the case of only three qubits in an external magnetic field. Nevertheless, we point out that quantum correlations cannot be induced solely by chaos and that these two are rather interconnected: chaos ensues when the entangled $W$ state becomes a ground state of the Heisenberg trimer.

Finally, we note that the Husimi structure, which we study here, is a good approximation to more realistic models with multisite interactions on ordinary lattices, being more reliable than the standard Bethe and mean-field approaches \cite{gujrati, pretti1}. Additionally, it can be used for the description of various polymers, and particularly, of RNA-like molecules \cite{rna}. Meanwhile, triangulated lattices, in their turn, are also of current great interest since they may describe real materials (e.g., copper based coordination compounds $\mathrm{Cu_9X_2(cpa)_6\cdot {\it n} H_2O}$; $\mathrm{X=F}$, $\mathrm{Cl}$, $\mathrm{Br}$ \cite{poli1, me22}), and additionally turn out to be an efficient ground for the understanding of specific mechanisms of quantum ordering at low temperatures \cite{YD}. Consequently, along with the aforementioned features, these arguments reinforce our interest in the studies of exactly solvable Ising-Heisenberg models on triangulated Husimi structures.

The paper is organized as follows: in Sec.~\ref{model} we introduce the multisite interaction spin-1/2 Ising-Heisenberg model on a THL and show its exact solution by means of the generalized star-triangle transformation and the recursive approach. In Sec.~\ref{bifurcation} we discuss the magnetic and entanglement properties of our model, and reveal their chaotic nature. Finally, we draw our conclusions in Sec.~\ref{concl}.

\section{Model}\label{model}

\begin{figure*}[ht]
\begin{center}
\small(a) \includegraphics[width=7.5cm]{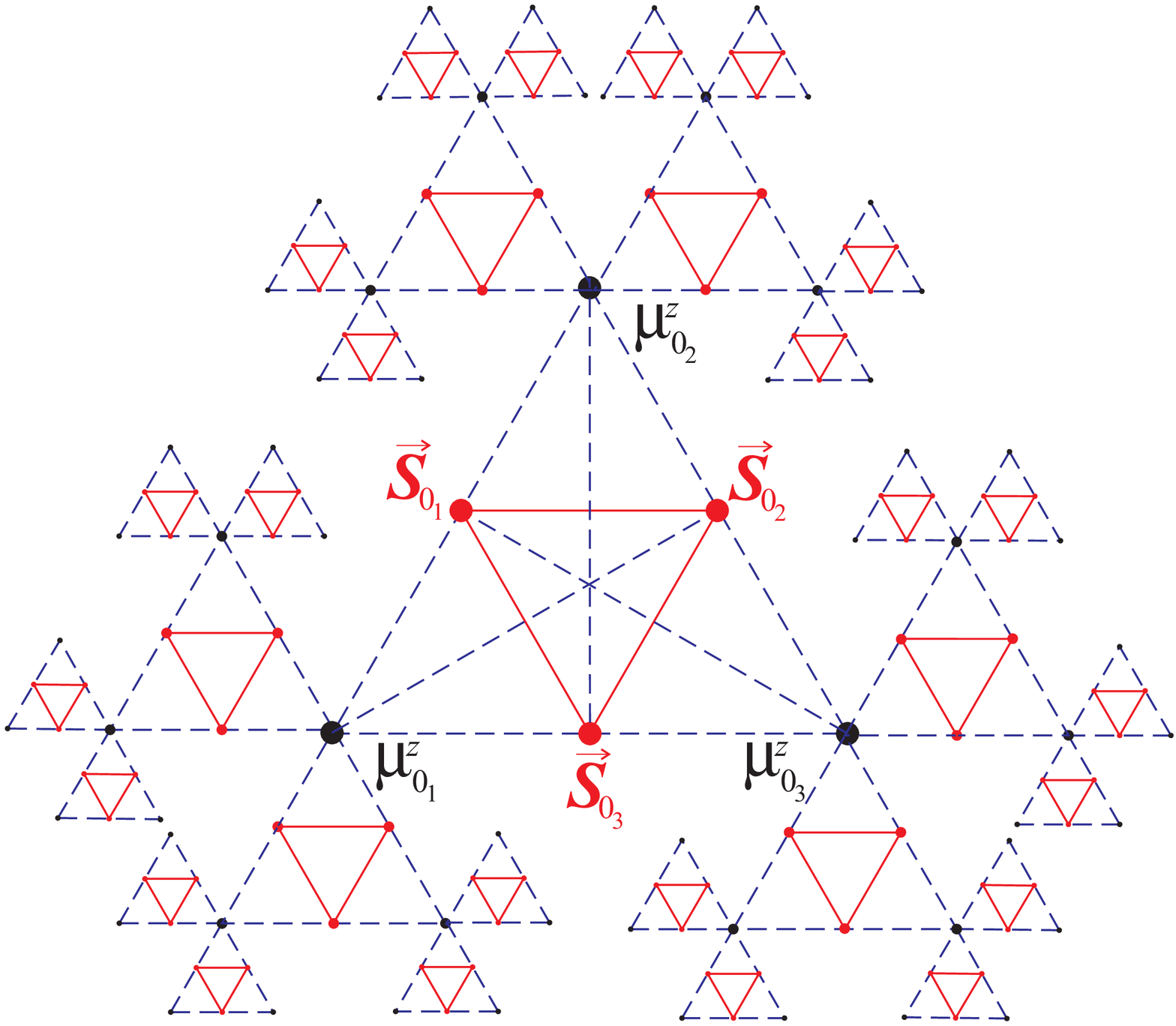}
\small(b) \includegraphics[width=7.5cm]{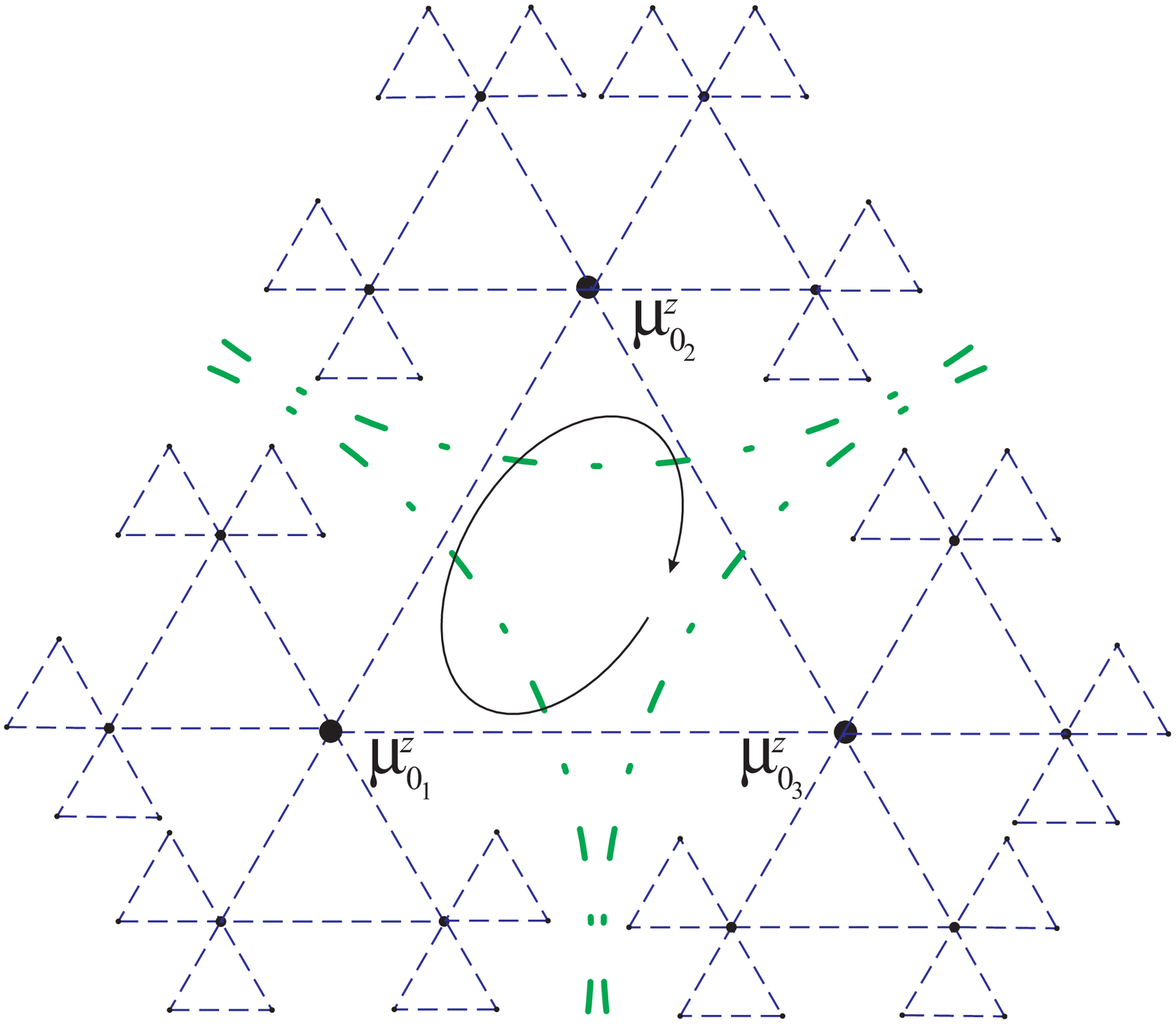}
\caption {(Color online) (a) The Ising-Heisenberg model on a triangulated Husimi lattice with coordination number $q=3$. The red (gray) circles denote the position of the Heisenberg $\vec{S}_{k_i}=\{S_{k_i}^x, S_{k_i}^y, S_{k_i}^z\}$ spins, and the black ones denote that of the Ising $\mu_{k_i}^z$ variables. The full red lines label the Heisenberg-type exchange interaction between $\vec{S}_{k_i}$ spins of the $k$th triangle, while the dashed blue ones stand for the Ising-type coupling of $S_{k_i}^z$ and $\mu_{k_j}^z$ ($i,j=1,2,3$); $S_{k_1}^z \mu_{k_{3}}^z$, $S_{k_2}^z \mu_{k_{1}}^z$ and $S_{k_3}^z \mu_{k_{2}}^z$ interactions are shown only for the central triangle. There is also a three-site exchange coupling $\mu_{k_1}^z\mu_{k_2}^z\mu_{k_3}^z$, that is not shown in this figure. (b) The effective Ising model on a Husimi lattice which the above Ising-Heisenberg model is mapped onto. The dashed blue lines label the two-site interactions, while the full black arrowed ellipse stands for the three-site interaction (shown only for the central triangle). The green dot-dashed lines show how the system is cut apart for employing the recursive method (see Sec.~\ref{recursive} for details). \label{lattice}}
\end{center}
\end{figure*}

We consider a recursive structure that is formed by a basic triangular unit bearing Ising variables on its vertices,
and Heisenberg-type spins situated on its bonds, which, in their turn, form an embedded triangle. Taking this building
block as the zeroth shell, we construct the next one, by connecting $q-1$ such units on each site of the bigger triangle,
and continue this process to develop higher-generation shells. Eventually we end up with a structure that can be called a
triangulated Husimi tree, since the Ising spins of this system constitute a regular Husimi tree with a coordination number $q$ [Fig.~\ref{lattice}(a)].
Note that the number of sites both at the surface and inside the tree grows exponentially, and for that reason we consider the
properties of the system in its deep interior (for more details on this see, e.g., Ref.~\cite{onsager}), i.e., on a triangulated Husimi lattice.

The definition of the spin-1/2 Ising-Heisenberg model on the above structure is now straightforward. The corresponding Hamiltonian reads:

\begin{eqnarray}\label{1}\nonumber
&&\mathcal{H}=\sum_{k=0}^{N_\triangle-1} \mathcal{H}_k, \\ \nonumber
&&\mathcal{H}_k=-J_3 \mu_{k_1}^z\mu_{k_2}^z\mu_{k_3}^z-\sum_{(i,j)}\left[J_{\mathrm{H}}^{xy}(S_{k_i}^x S_{k_j}^x+S_{k_i}^y S_{k_j}^y)+\right. \\ \nonumber
&&\left.J_{\mathrm{H}}^{zz}S_{k_i}^z S_{k_j}^z\right]-J_{\mathrm{IH}}(\mu_{k_1}^z+\mu_{k_2}^z+\mu_{k_3}^z)(S_{k_1}^z+S_{k_2}^z+S_{k_3}^z)- \\
&&\frac{H_\mathrm{I}}{q}(\mu_{k_1}^z+\mu_{k_2}^z+\mu_{k_3}^z)-H_\mathrm{H} (S_{k_1}^z+S_{k_2}^z+S_{k_3}^z).
\end{eqnarray}
Here $\mathcal{H}_k$ denotes the Hamiltonian of the $k$th triangle ($N_\triangle$ being the number of these triangles), $\vec{S}_{k_i}=\{S_{k_i}^x, S_{k_i}^y, S_{k_i}^z\}$ stands for Heisenberg spin-1/2 operators, while $\mu_{k_i}^z$ for that of the Ising ones ($i=1, 2, 3$). The first term in $\mathcal{H}_k$ corresponds to the three-site interaction of a strength $J_3$ between nodal Ising spins, $J_{\mathrm{IH}}$ is the strength of the Ising-to-Heisenberg coupling, whereas $J_{\mathrm{H}}^{xy}$ and $J_{\mathrm{H}}^{zz}$ are the interaction strengths of the Heisenberg exchange coupling on the $(x,y)$ plane and the $z$ direction, respectively  (equivalently, this interaction can be defined as of a strength $J\equiv J_{\mathrm{H}}^{zz}$, with the anisotropy parameter $\Delta\equiv J_{\mathrm{H}}^{xy}/J_{\mathrm{H}}^{zz}$). The system is also subject to external magnetic fields $H_\mathrm{I}$ (note that each Ising spin belongs simultaneously to $q$ triangles) and $H_\mathrm{H}$, acting upon the Ising- and Heisenberg-type variables, and directed along the $z$ axis [see also Fig.~\ref{lattice}(a)].

%As already mentioned in Sec.~\ref{intr}, due to the inclusion of three-site interactions as well as of the external magnetic field,
%this model is rather different from the one discussed in Ref.~\cite{strecka}, and exhibits qualitatively different behavior, as shown below.

\subsection{The generalized star-triangle transformation and the effective Hamiltonian}\label{effective}

To solve the above defined model exactly we employ the generalized star-triangle ($Y-\Delta$) transformation \cite{YD, YD1, strecka}, that allows one to map the initial Ising-Heisenberg model on a THL onto an effective Ising model on a Husimi lattice. For that, owing to the fact that the triangular Hamiltonians $\mathcal{H}_k$ are commutative with one another, i.e., $[\mathcal{H}_k, \mathcal{H}_l]=0$, we partially factorize the partition function of the initial system to partition functions of triangular Hamiltonians (we work in units such that the Boltzmann constant $k_B$ is unity):
\begin{eqnarray}\label{2}
\begin{split}
&\mathcal{Z}=\sum_{\{\mu_{k_i}\}}\prod_{k=0}^{N_\triangle-1}\mathrm{Tr}_k \exp(-\mathcal{H}_k/T)=\\
&\sum_{\{\mu_{k_i}\}}\prod_{k=0}^{N_\triangle-1}\mathcal{Z}_k(\mu_{k_1}^z, \mu_{k_2}^z, \mu_{k_3}^z),
\end{split}
\end{eqnarray}
Here $T$ is the temperature, $\mathrm{Tr}_k$ denotes the trace over the degrees of freedom of the three Heisenberg spins of the $k$th triangle, whereas the subsequent summation runs over all the possible configurations of the nodal Ising spins, $\mathcal{Z}_k(\mu_{k_1}^z, \mu_{k_2}^z, \mu_{k_3}^z)$ having the form
\begin{eqnarray}\label{3}
&&\mathcal{Z}_k(\mu_{k_1}^z, \mu_{k_2}^z, \mu_{k_3}^z)=\\ \nonumber
&&\exp \left[\frac{J_3}{T}\mu_{k_1}^z\mu_{k_2}^z\mu_{k_3}^z+\frac{H_\mathrm{I}}{qT}(\mu_{k_1}^z+\mu_{k_2}^z+\mu_{k_3}^z)\right] W(\mu_{k}).
\end{eqnarray}
In this equation $\mu_k=\pm3/2, \pm1/2$ are the eigenvalues of the total Ising spin operator of the $k$th triangle $\mu_k^z=\mu_{k_1}^z+\mu_{k_2}^z+\mu_{k_3}^z$ of a specific spin configuration, and $W(\mu_{k})$ is given as:
\begin{eqnarray}\label{4}\nonumber
&&W(\mu_{k})=2\cosh \left(\frac{H_\mathrm{H}+J_{\mathrm{IH}} \mu _{k}}{2T}\right)\times \\ \nonumber
&&\left[\exp\left(\frac{J_\text{H}^{xy}}{T}-\frac{J_\text{H}^{zz}}{4T}\right) +2\exp\left(-\frac{J_\text{H}^{xy}}{2T} - \frac{J_\text{H}^{zz}}{4 T}\right)\right]+ \\
&&2\cosh \left[\frac{3 \left(H_\mathrm{H}+J_{\mathrm{IH}}\mu _{k}\right)}{2T}\right] \exp\left(\frac{3 J_\text{H}^{zz}}{4 T}\right).
\end{eqnarray}
Having brought the partition function to the form of Eq.~(\ref{3}), we apply the generalized star-triangle transformation, for mapping the initial Ising-Heisenberg model onto an effective Ising one:
\begin{widetext}
\begin{eqnarray}\label{5}\nonumber
&&\mathcal{Z}(\mu_{k_1}^z, \mu_{k_2}^z, \mu_{k_3}^z)=\exp \left[\frac{J_3}{T}\mu_{k_1}^z\mu_{k_2}^z\mu_{k_3}^z+\frac{H_\mathrm{I}}{qT}(\mu_{k_1}^z+\mu_{k_2}^z+\mu_{k_3}^z)\right] W(\mu_{k})= \\
&&A\exp\left[\frac{J_3^{\text{eff}}}{T}\mu_{k_1}^z\mu_{k_2}^z\mu_{k_3}^z+\frac{J_2^{\text{eff}}}{T}\left(\mu_{k_1}^z\mu_{k_2}^z+\mu_{k_2}^z\mu_{k_3}^z+\mu_{k_1}^z\mu_{k_3}^z\right)+\frac{H^{\text{eff}}}{qT}(\mu_{k_1}^z+\mu_{k_2}^z+\mu_{k_3}^z)\right].
\end{eqnarray}
\end{widetext}
As the above expression should hold true for any configuration of Ising spins, it results in a set of four equations (the initial eight ones corresponding to $2^3=8$ configurations of the three nodal spins are reduced to just four due to a threefold degeneracy of $\mu_k=1/2$ and $\mu_k=-1/2$ states). Eventually, these equations determine unambiguously the mapping parameters through the following expressions:
\begin{eqnarray}\label{6}\nonumber
&&A=[W(3/2)W(-3/2)]^{1/8}[W(1/2)W(-1/2)]^{3/8},\\
&&J_3^{\text{eff}}=J_3+T\ln\left( \frac{W(3/2)}{W(-3/2)}\cdot\left[\frac{W(-1/2)}{W(1/2)}\right]^{3}\right), \\ \nonumber
&&J_2^{\text{eff}}=\frac{T}{2}\ln W(3/2)W(-3/2)}{W(1/2)W(-1/2), \\ \nonumber
&&H^{\text{eff}}=H_\mathrm{I}+\frac{qT}{4}\ln \frac{W(3/2)W(1/2)}{W(-1/2)W(-3/2)}.
\end{eqnarray}
Consequently, owing to Eq.~(\ref{3}) we rewrite the partition function of the initial Hamiltonian of Eq.~(\ref{1}) as
\begin{eqnarray}\label{7}
\begin{split}
&\mathcal{Z}(J_3, J_\mathrm{H}^{xy}, J_\mathrm{H}^{zz}, J_{\mathrm{IH}}, H_\mathrm{I}, H_\mathrm{H}, T, q)=\\
&A^{N_{\triangle}}\mathcal{Z}^{\text{eff}}(J_3^{\text{eff}}, J_2^{\text{eff}}, H^{\text{eff}}, T, q),
\end{split}
\end{eqnarray}
which maps the spin-1/2 Ising-Heisenberg model on a THL onto a spin-1/2 Ising model on a simple Husimi lattice with effective three- and two-site interactions in an effective magnetic field. The corresponding Hamiltonian reads [see also Fig.~\ref{lattice}(b)]:
\begin{widetext}
\begin{eqnarray} \label{8}
\mathcal{H}^{\text{eff}}=-\sum_{k=0}^{N_\triangle-1}\left[J_3^{\text{eff}} \mu_{k_1}^z\mu_{k_2}^z\mu_{k_3}^z+J_2^{\text{eff}}\left(\mu_{k_1}^z\mu_{k_2}^z+\mu_{k_2}^z\mu_{k_3}^z+\mu_{k_1}^z\mu_{k_3}^z\right)+\frac{H^{\text{eff}}}{q}(\mu_{k_1}^z+\mu_{k_2}^z+\mu_{k_3}^z)\right].
\end{eqnarray}
\end{widetext}
The distinguishable feature of this final model is the fact that it can be solved exactly through the recursive method, that we present in the next subsection. We also make a note here that for the values of parameters that we consider below, the effective coupling $J_2^\text{eff}$ always remains ferromagnetic.

Meanwhile, we express all the physical quantities of interest in terms of the above defined effective parameters. Specifically, denoting by $f_\triangle$ and $f_\triangle^{\text{eff}}$ the free energy per triangle of the initial and effective models, from Eq.~(\ref{7}) we find:
\begin{equation}
f_\triangle=f_\triangle^{\text{eff}}-T\ln A. \label{9}
\end{equation}
Consequently, the single-site magnetization $m_\mathrm{I}$ of the nodal Ising spins reads (there are $3/q$ of them in each triangle):
\begin{equation} \label{10}
m_\mathrm{I}=-\frac{q}{3} \frac{\partial f_\triangle}{\partial H_\mathrm{I}}=-\frac{q}{3} \frac{\partial f_\triangle^{\text{eff}}}{\partial H^{\text{eff}}}\frac{\partial H^{\text{eff}}}{\partial H_\mathrm{I}}=m^{\text{eff}}.
\end{equation}
In other words, $m_\mathrm{I}$ is equal to the per-cite magnetization $m^{\text{eff}}$ of the effective model, which is, however, not the case for the single-site magnetization $m_\mathrm{H}$ of the Heisenberg sublattice:
\begin{eqnarray}\label{11}\nonumber
&&m_\mathrm{H}=-\frac{1}{3} \frac{\partial f_\triangle}{\partial H_\mathrm{H}}=\\ \nonumber
&&-\frac{1}{3} \left(\frac{\partial f_\triangle^{\text{eff}}}{\partial H^{\text{eff}}}\frac{\partial H^{\text{eff}}}{\partial H_\mathrm{H}}+
\frac{\partial f_\triangle^{\text{eff}}}{\partial J_3^{\text{eff}}}\frac{\partial J_3^{\text{eff}}}{\partial H_\mathrm{H}}+
\frac{\partial f_\triangle^{\text{eff}}}{\partial J_2^{\text{eff}}}\frac{\partial J_2^{\text{eff}}}{\partial H_\mathrm{H}}\right)+ \\ \nonumber
&&\frac{T}{3}\frac{\partial \ln A}{\partial H_\mathrm{H}}=\frac{m^{\text{eff}}}{q}\frac{\partial H^{\text{eff}}}{\partial H_\mathrm{H}}+
\frac{\varepsilon_3^{\text{eff}}}{3}\frac{\partial J_3^{\text{eff}}}{\partial H_\mathrm{H}}+\frac{\varepsilon_2^{\text{eff}}}{3}\frac{\partial J_2^{\text{eff}}}{\partial H_\mathrm{H}}+\\
&&\frac{T}{3A}\frac{\partial A}{\partial H_\mathrm{H}},
\end{eqnarray}
where $\varepsilon_3^{\text{eff}}=\langle \mu_{k_1}^z\mu_{k_2}^z\mu_{k_3}^z \rangle$ and $\varepsilon_2^{\text{eff}}=\langle \mu_{k_1}^z\mu_{k_2}^z+\mu_{k_2}^z\mu_{k_3}^z+\mu_{k_1}^z\mu_{k_3}^z \rangle$ are the three- and two-site spin-spin correlation functions of the effective Ising model.

In what follows we are also interested in the nearest-neighbor two-site correlation functions of the Heisenberg spins. Note that as the magnetic field is directed along the $z$ axis, the correlations in the $x$ and $y$ directions are equal. Therefore, the per-pair spin-spin correlation functions read:
\begin{eqnarray}\label{12}\nonumber
&&\varepsilon_{xx}=\varepsilon_{yy}=-\frac{1}{3} \frac{\partial f_\triangle}{\partial J_{xy}}=\\ \nonumber
&&-\frac{1}{3} \left(\frac{\partial f_\triangle^{\text{eff}}}{\partial H^{\text{eff}}}\frac{\partial H^{\text{eff}}}{\partial J_{xy}}+
\frac{\partial f_\triangle^{\text{eff}}}{\partial J_3^{\text{eff}}}\frac{\partial J_3^{\text{eff}}}{\partial J_{xy}}+
\frac{\partial f_\triangle^{\text{eff}}}{\partial J_2^{\text{eff}}}\frac{\partial J_2^{\text{eff}}}{\partial J_{xy}}\right)+\\ \nonumber
&&\frac{T}{3}\frac{\partial \ln A}{\partial J_{xy}}=\frac{m^{\text{eff}}}{q}\frac{\partial H^{\text{eff}}}{\partial J_{xy}}+
\frac{\varepsilon_3^{\text{eff}}}{3}\frac{\partial J_3^{\text{eff}}}{\partial J_{xy}}+\frac{\varepsilon_2^{\text{eff}}}{3}\frac{\partial J_2^{\text{eff}}}{\partial J_{xy}}+ \\
&&\frac{T}{3A}\frac{\partial A}{\partial J_{xy}}.
\end{eqnarray}
Similarly, the two-site spin-spin correlation function in the $z$ direction is given as:

\begin{eqnarray}\label{13}\nonumber
&&\varepsilon_{zz}=-\frac{1}{3} \frac{\partial f_\triangle}{\partial J_{zz}}=\\ \nonumber
&&-\frac{1}{3} \left(\frac{\partial f_\triangle^{\text{eff}}}{\partial H^{\text{eff}}}\frac{\partial H^{\text{eff}}}{\partial J_{zz}}+
\frac{\partial f_\triangle^{\text{eff}}}{\partial J_3^{\text{eff}}}\frac{\partial J_3^{\text{eff}}}{\partial J_{zz}}+
\frac{\partial f_\triangle^{\text{eff}}}{\partial J_2^{\text{eff}}}\frac{\partial J_2^{\text{eff}}}{\partial J_{zz}}\right)+\\ \nonumber
&&\frac{T}{3}\frac{\partial \ln A}{\partial J_{zz}}=\frac{m^{\text{eff}}}{q}\frac{\partial H^{\text{eff}}}{\partial J_{zz}}+
\frac{\varepsilon_3^{\text{eff}}}{3}\frac{\partial J_3^{\text{eff}}}{\partial J_{zz}}+\frac{\varepsilon_2^{\text{eff}}}{3}\frac{\partial J_2^{\text{eff}}}{\partial J_{zz}}+ \\
&&\frac{T}{3A}\frac{\partial A}{\partial J_{zz}}.
\end{eqnarray}

\subsection{Exact solution of the effective model by means of the recursive method}\label{recursive}

As mentioned previously, the effective Ising model on a Husimi lattice, defined by the Hamiltonian (\ref{8}), can be solved exactly through the recursive method. The latter takes into account the specific $-$ recursive $-$ structure of the lattice, that can be cut apart at the central unit (the central triangle for the case of a Husimi lattice), being divided into $q$ identical branches [Fig~\ref{lattice}(b)]. Consequently, the partition function is expressed in terms of the partition functions of each of those branches  \cite{pretti, monroe, anan, comm, jetp}:
\begin{eqnarray}\label{14}\nonumber
&&\mathcal{Z}_\text{eff}=\sum_{\mu_{0_1}, \mu_{0_2},\mu_{0_3}}\exp\left[\frac{J_3^{\text{eff}}}{T}\mu_{0_1}\mu_{0_2}\mu_{0_3}+\right.\\
&&\frac{J_2^{\text{eff}}}{T}\left(\mu_{0_1}\mu_{0_2}+\mu_{0_2}\mu_{0_3}+\mu_{0_1}\mu_{0_3}\right)+\\ \nonumber
&&\left.\frac{H^\text{eff}}{qT}(\mu_{0_1}+\mu_{0_2}+\mu_{0_3})\right] g_n^{q-1}(\mu_{0_1})  g_n^{q-1}(\mu_{0_2}) g_n^{q-1}(\mu_{0_3}),
\end{eqnarray}
where the summation runs over the eigenvalues $\mu_{0_i}=\pm 1/2$ of the Ising  $\mu_{0_i}^z$ spin operators (i.e., over the possible spin configurations), $g_n(\mu_{0_i})$ standing for the contribution to the partition function of each of the $q-1$ branches that contain $n$ shells. Next, we repeat the cutting-apart procedure once again, and e.g., for $g_n(\mu_{0_1})$ we find:
\begin{eqnarray}\label{15}\nonumber
&& g_n(\mu_{0_1})=\sum_{\mu_{1_2},\mu_{1_3}}\exp\left[\frac{J_3^{\text{eff}}}{T}\mu_{0_1}\mu_{1_2}\mu_{1_3}+\right.\\
&&\frac{J_2^{\text{eff}}}{T}\left(\mu_{0_1}\mu_{1_2}+\mu_{1_2}\mu_{1_3}+\mu_{0_1}\mu_{1_3}\right)+\\ \nonumber
&&\left.\frac{H^\text{eff}}{qT}(\mu_{0_1}+\mu_{1_2}+\mu_{1_3})\right]g_{n-1}^{q-1}(\mu_{1_2}) g_{n-1}^{q-1}(\mu_{1_3}).
\end{eqnarray}
Eventually, this procedure results in a recursive relation for $x_n=g_n(1/2)/g_n(-1/2)$ (see, e.g. Refs.~\cite{pre, comm, jetp}):
\begin{widetext}
\begin{eqnarray}\label{16}
\begin{split}
&x_n=f(x_{n-1}), \\
&f(x)=\frac{x^{2(q-1)}\exp\left(\frac{J_3^{\text{eff}}}{4T}+\frac{3H^{\text{eff}}}{qT}\right)+2 x^{q-1}\exp\left({-\frac{J_2^{\text{eff}}}{T}+\frac{2H^{\text{eff}}}{qT}}\right)+\exp\left({\frac{J_3^{\text{eff}}}{4T}-\frac{J_2^{\text{eff}}}{T}}+\frac{H^{\text{eff}}}{qT}\right)}
{x^{2(q-1)}\exp\left(-\frac{J_2^{\text{eff}}}{T}+\frac{2H^{\text{eff}}}{qT}\right) +2x^{q-1} \exp\left(\frac{J_3^{\text{eff}}}{4T}-\frac{J_2^{\text{eff}}}{T}+\frac{H^{\text{eff}}}{qT}\right)+1}.
\end{split}
\end{eqnarray}
\end{widetext}
Although $x_n$ does not have a direct physical meaning, the mapping $f(x)$ describes the properties of the system. Specifically, the above defined single-site magnetization $m_{\text{I}}$, as well as the three- and two-site correlation functions $\varepsilon^{\text{eff}}_3$ and $\varepsilon^{\text{eff}}_2$ of spins situated deep inside the tree (where the surface effects are negligible), can be expressed straightforwardly in terms of $x_n$:
\begin{widetext}
\begin{eqnarray}\label{17}
\begin{split}
&m^{\text{eff}}=\frac{1}{2}\frac{x_n^q e^{\frac{H^{\text{\text{eff}}}}{q T}}-1}{x_n^qe^{\frac{H^{\text{\text{eff}}}}{q T}}+1}, \\
&\varepsilon_3^{\text{eff}}=\frac{1}{8}\frac{x_n^{3(q-1)}\exp\left(\frac{J_3^{\text{eff}}}{4T}+\frac{3H^{\text{eff}}}{qT}\right)-3 x_n^{2(q-1)}\exp\left({-\frac{J_2^{\text{eff}}}{T}+\frac{2H^{\text{eff}}}{qT}}\right)+3x_n^{q-1}\exp\left(\frac{J_3^{\text{eff}}}{4T}-\frac{J_2^{\text{eff}}}{T}+\frac{H^{\text{eff}}}{qT}\right)-1}
{x_n^{3(q-1)}\exp\left(\frac{J_3^{\text{eff}}}{4T}+\frac{3H^{\text{eff}}}{qT}\right)+3 x_n^{2(q-1)}\exp\left({-\frac{J_2^{\text{eff}}}{T}+\frac{2H^{\text{eff}}}{qT}}\right)+
3x_n^{q-1}\exp\left(\frac{J_3^{\text{eff}}}{4T}-\frac{J_2^{\text{eff}}}{T}+\frac{H^{\text{eff}}}{qT}\right)+1}, \\
&\varepsilon_2^{\text{eff}}=\frac{1}{4}\frac{3x_n^{3(q-1)}\exp\left(\frac{J_3^{\text{eff}}}{4T}+\frac{3H^{\text{eff}}}{qT}\right)- 3x_n^{2(q-1)}\exp\left({-\frac{J_2^{\text{eff}}}{T}+\frac{2H^{\text{eff}}}{qT}}\right)-3x_n^{q-1}\exp\left(\frac{J_3^{\text{eff}}}{4T}-\frac{J_2^{\text{eff}}}{T}+\frac{H^{\text{eff}}}{qT}\right)+3}
{x_n^{3(q-1)}\exp\left(\frac{J_3^{\text{eff}}}{4T}+\frac{3H^{\text{eff}}}{qT}\right)+3 x_n^{2(q-1)}\exp\left({-\frac{J_2^{\text{eff}}}{T}+\frac{2H^{\text{eff}}}{qT}}\right)+
3x_n^{q-1}\exp\left(\frac{J_3^{\text{eff}}}{4T}-\frac{J_2^{\text{eff}}}{T}+\frac{H^{\text{eff}}}{qT}\right)+1}.
\end{split}
\end{eqnarray}
\end{widetext}
In other words, in the limit $n\rightarrow \infty$ the mapping $f(x)$ determines the state of the system, and particularly, the canonical ensemble averages of all physical quantities that we are interested in here. Consequently, Eqs.~(\ref{10})$-$(\ref{13}) along with Eqs.~(\ref{16}) and (\ref{17}) provide the exact solution of the spin-1/2 Ising-Heisenberg model on a THL.

\subsection{Pairwise entanglement of the Heisenberg trimer}\label{entangl}

In this subsection we turn to characterizing the entanglement properties of the Heisenberg trimers of the THL. To quantify the amount of quantum correlations that the Heisenberg spins of each triangle share, we use below the concurrence \cite{wooters}, as a computable entanglement measure of a bipartite mixed state. For this purpose we require the reduced density matrix of a pair of Heisenberg spins (i.e., qubits), taking into account that due to the cyclic symmetry that each Heisenberg triangle possesses, the entanglement of any of its pairs is identical. We also note that as the neighboring Heisenberg triangles are coupled to one another by means of Ising-type (diagonal) interaction, they are not entangled with each other.

In order to express the reduced density matrix of a pair of triangular qubits (that we denote by $\rho_{12}$), in terms of the above defined magnetization $m_{\text{H}}$ and the spin-spin correlation functions $\varepsilon_{xx}$, $\varepsilon_{yy}$ and $\varepsilon_{zz}$, we right it down in the Hilbert-Schmidt basis \cite{hillbert}:
\begin{eqnarray}\label{18}\nonumber
&&\rho_{12}=\frac{1}{4}\left(I\otimes I+2\sum_{i=1}^3\left(a_i\sigma_i\otimes I+b_i I\otimes \sigma_i \right)+\right.\\
&&\left.\sum_{i,j=1}^3t_{ij}\sigma_i\otimes\sigma_j\right).
\end{eqnarray}
Here $I$ is a $2\times2$ identity matrix, and $\sigma_i$'s are the standard Pauli matrices, related to the spin operators as $S^x=\frac{1}{2}\sigma_1$, $S^y=\frac{1}{2}\sigma_2$ $S^z=\frac{1}{2}\sigma_3$. Following the well-known procedure we express the coefficients of the above decomposition through the single-site magnetization and spin-spin correlation functions:
\begin{eqnarray}\label{19}\nonumber
&&a_3=b_3=m_{\text{H}}, \,\, t_{11}=4\varepsilon_{xx}, \\
&&t_{22}=4\varepsilon_{yy}, \,\, t_{33}=4\varepsilon_{zz},
\end{eqnarray}
whereas all the other coefficients are equal to zero. As a result, the reduced density matrix of a pair of qubits reads:
\begin{eqnarray}\label{20}
&&\rho_{{12}}=\\
&&\left(
\begin{array}{llll}
 \frac{1}{4}+\varepsilon_{zz}+m_{\text{H}} & 0 & 0 & 0 \\
 0 & \frac{1}{4}-\varepsilon_{zz} & 2\varepsilon_{xx} & 0 \\
 0 & 2\varepsilon_{xx} & \frac{1}{4}-\varepsilon_{zz} & 0 \\
 0 & 0 & 0 & \frac{1}{4}+\varepsilon_{zz}-m_{\text{H}} \nonumber
\end{array}
\right),
\end{eqnarray}
that is a special case of the so-called $X$ state \cite{xstate}. Its concurrence, as known, takes the following form:
\begin{eqnarray}\label{20}
&&C(\rho_{12})=\\ \nonumber
&&2\max\left(2|\varepsilon_{xx}|-\sqrt{\left(\frac{1}{4}+\varepsilon_{zz}-m_{\text{H}}\right) \left(\frac{1}{4}+\varepsilon_{zz}+m_{\text{H}}\right)}, 0\right).
\end{eqnarray}
Note that this final expression for the pairwise entanglement of the triangulated Husimi lattice is exact and takes into account the specific recursive structure of the system.

\section{Bifurcation and chaotic behavior of the magnetization and entanglement}\label{bifurcation}

Having the exact solution of the spin-1/2 Ising-Heisenberg model on a THL we proceed to the discussion of its magnetic and entanglement properties.

In what follows we are mainly interested in the case of isotropic antiferromagnetic Heisenberg interactions, i.e., $J_\text{H}^{xy}=J_\text{H}^{zz}\equiv J_\text{H}<0$, and the three-site interaction $J_3$ is also considered to be of antiferromagnetic character ($J_3<0$). Nevertheless, we choose the Heisenberg interactions to be dominant ($J_3/J_\text{H}<1$), which reinforces our assumption of having an alternating sequence of Ising and Heisenberg variables (for weaker interactions the fluctuations in the transverse direction are expected to be reduced in the presence of a magnetic field directed along the $z$ axis, which may give rise to spin-spin coupling of Ising, rather than of Heisenberg character). Additionally, we assume the ferromagnetic Ising-to-Heisenberg coupling to be weak, by fixing the corresponding ratio to $J_\text{IH}/J_\text{H}=-0.01$ (as already mentioned, the effective two-site Ising interaction $J_2^\text{eff}$ always remains ferromagnetic here).  Furthermore, we set the magnetic fields acting upon the Ising and Heisenberg spins equal: $H_\text{H}=H_\text{I}\equiv H$. Finally, we note that the initial [Eq.~(\ref{1})], as well as the effective [Eq.~(\ref{8})] systems possess a particular symmetry, namely, their properties remain unchanged under the replacement $J_3\rightarrow -J_3$ and $H \rightarrow-H$ (below we consider the region of positive magnetic fields).

\subsection{Magnetization}\label{magnb}

\begin{figure}[ht]
\begin{center}
\small(a) \includegraphics[width=7cm]{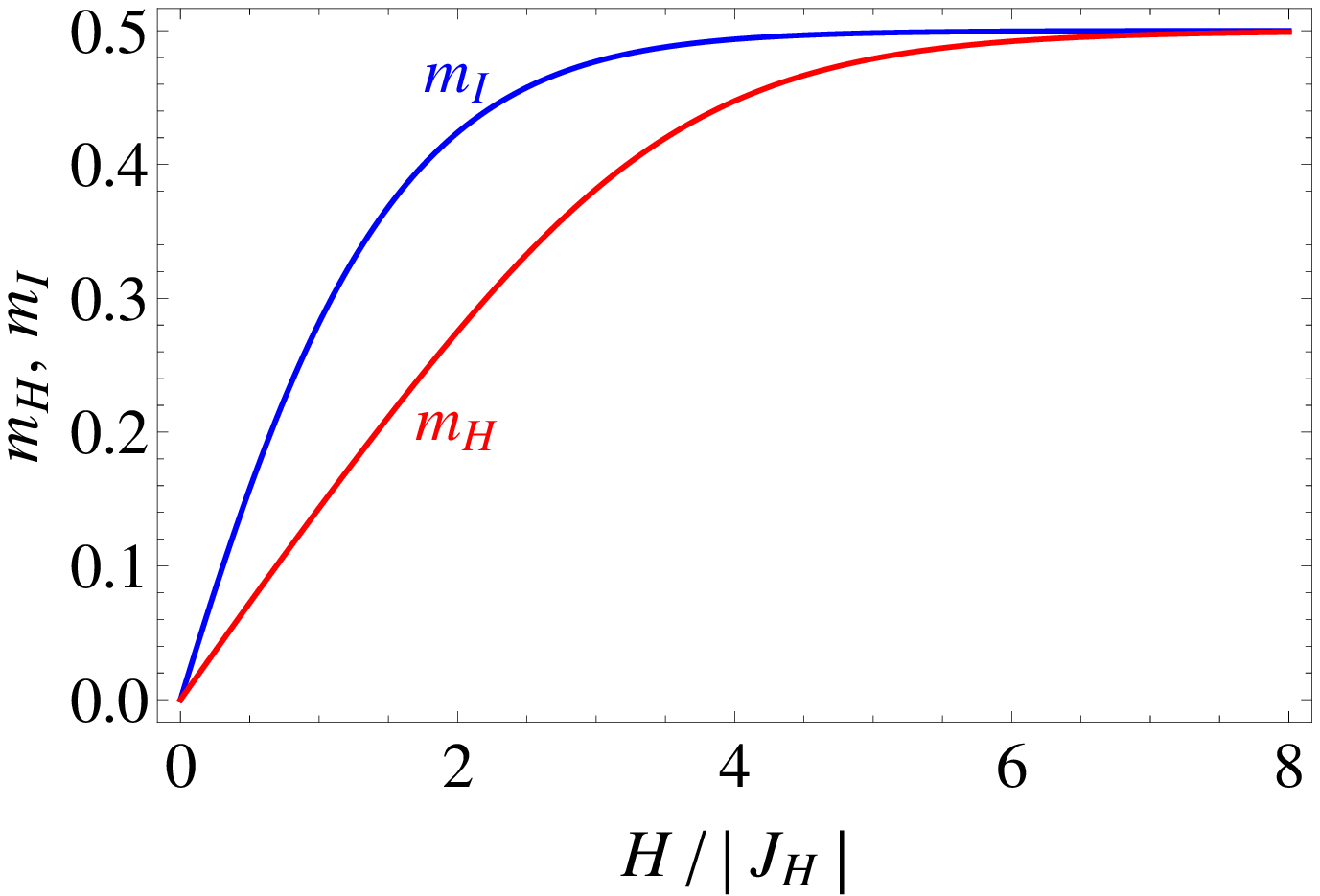}\\
\small(b) \includegraphics[width=7cm]{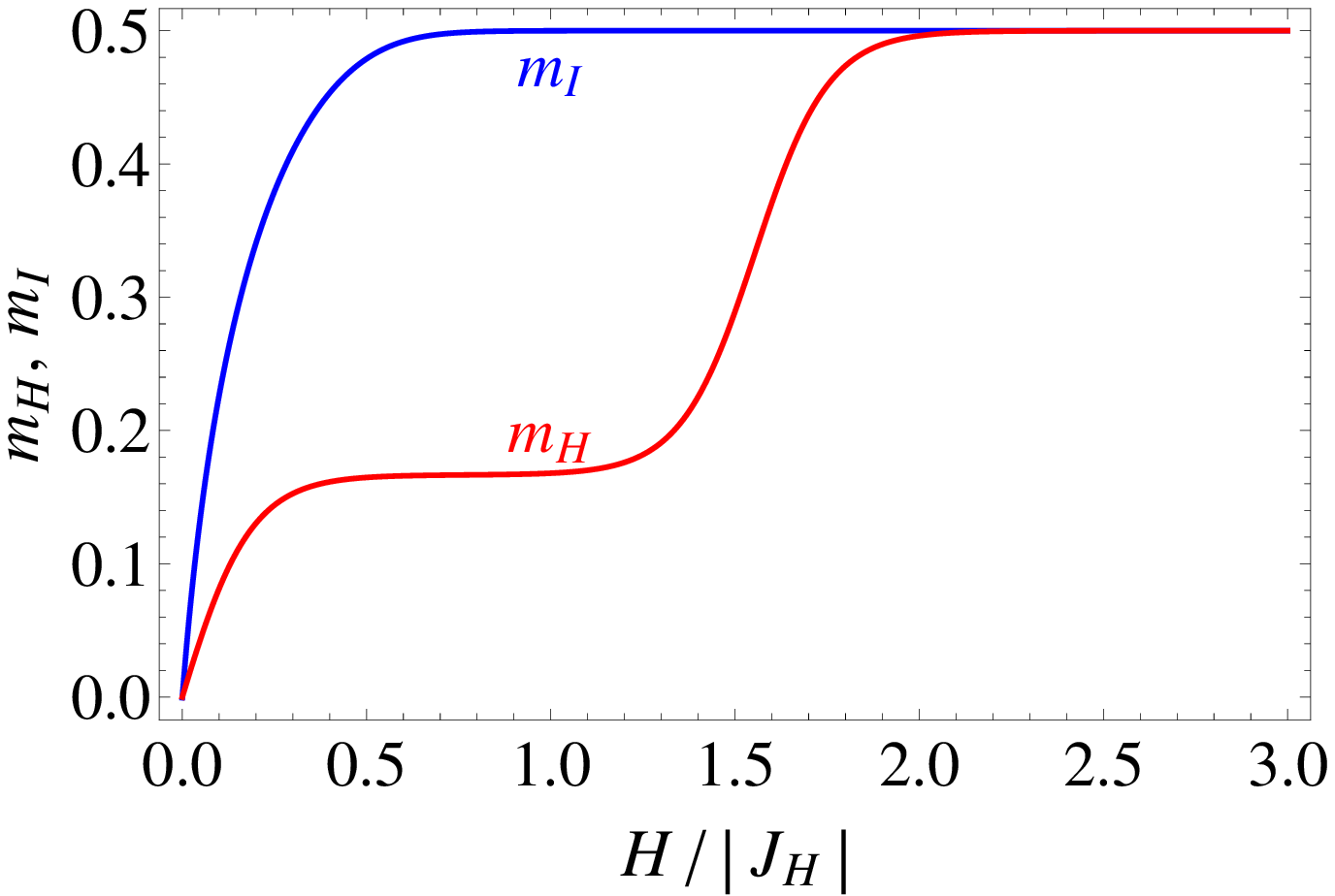}
\caption {(Color online) (a)  The single-site magnetization $m_{\text{H}}$ of the Heisenberg (red curve) and $m_{\text{I}}$ of the Ising (blue curve) sublattices versus the magnetic field $H_\text{H}=H_\text{I}\equiv H$. Here $J_{\text{IH}}/J_{\text{H}}=-0.01$, $J_3/J_\text{H}=0.5$, $T/|J_\text{H}|=1$ and $q=3$ ($J_{\text{H}}^{xy}=J_{\text{H}}^{zz}\equiv J_{\text{H}}$); (b) The same, but for $T/|J_\text{H}|=0.1$. \label{mhigh}}
\end{center}
\end{figure}

We start with the magnetic properties at relatively high temperatures, where the system possesses usual properties as shown in Fig.~\ref{mhigh}(a) $-$ both for Ising- and Heisenberg-type spins the magnetization goes monotonically to saturation with the increase in the strength of the magnetic field. Nevertheless, as the temperature is decreased we find the appearance of magnetization plateaus of the Heisenberg trimer at 1/6 of the saturation value, that corresponds to the ground states $|\psi\mathrm\rangle=\frac{1}{\sqrt{3}}\left(|\uparrow\uparrow\downarrow\rangle+|\uparrow\downarrow\uparrow\rangle+|\downarrow\uparrow\uparrow\rangle\right)$ for $H>0$ and $|\psi\mathrm\rangle=\frac{1}{\sqrt{3}}\left(|\downarrow\downarrow\uparrow\rangle+|\downarrow\uparrow\downarrow\rangle+|\uparrow\downarrow\downarrow\rangle\right)$ for $H<0$ (note that these states are highly entangled $W$ states to which we return in the next subsection).

\begin{figure}[ht]
\begin{center}
\small(a) \includegraphics[width=7cm]{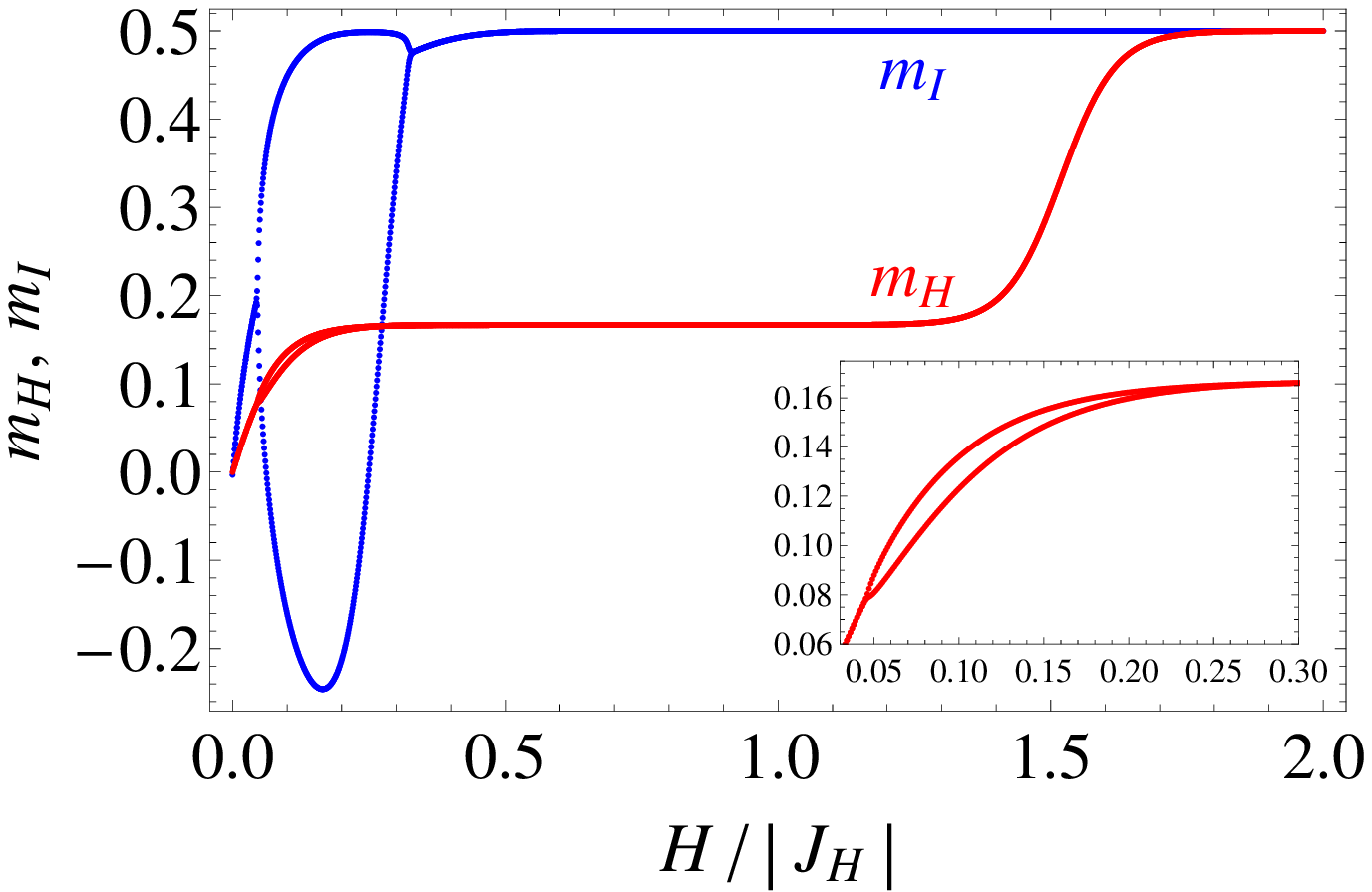}\\
\small(b) \includegraphics[width=7cm]{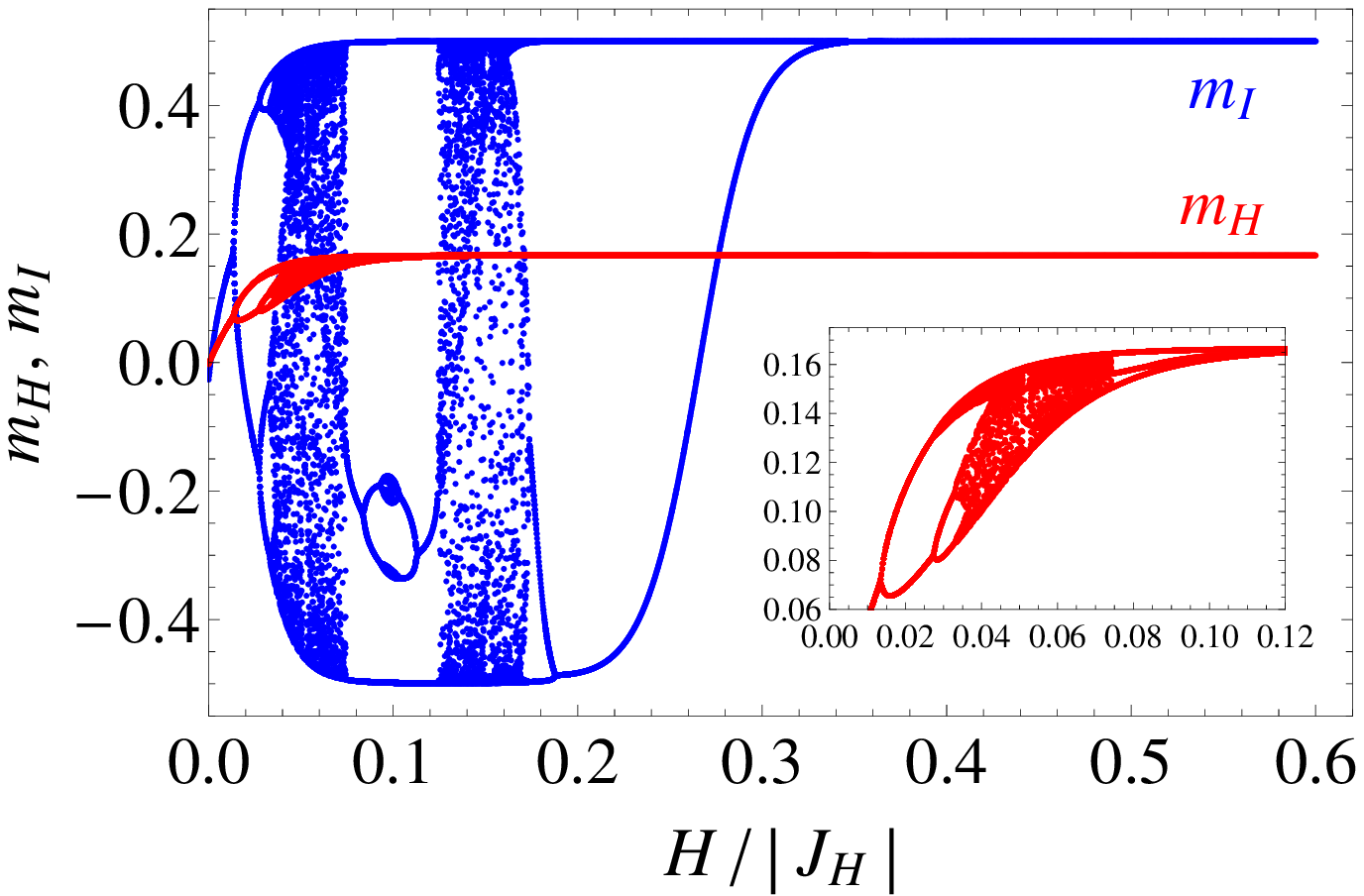}
\caption {(Color online) (a) The single-site magnetization $m_{\text{H}}$ of the Heisenberg (red curve) and $m_{\text{I}}$ of the Ising (blue curve) sublattices versus the magnetic field $H_\text{H}=H_\text{I}\equiv H$. Here $J_{\text{IH}}/J_{\text{H}}=-0.01$, $J_3/J_\text{H}=0.5$, $T/|J_\text{H}|=0.05$ and $q=3$ ($J_{\text{H}}^{xy}=J_{\text{H}}^{zz}\equiv J_{\text{H}}$). The inset shows the details in the area of a period doubling of the magnetization $m_{\text{H}}$; (b) The same, but for $T/|J_\text{H}|=0.02$ (the inset shows the details of the chaotic behavior of the magnetization $m_{\text{H}}$). \label{mlow}}
\end{center}
\end{figure}

The above picture, is, however, well known, and we do not go into more details in this respect. Instead, we consider the properties of the system at even lower temperatures, where, due to the specific lattice structure, it exhibits a doubling bifurcation $-$ the mapping $f(x)$ instead of a stable fixed  point converges now to a stable two-periodic cycle. Consequently, the Ising $m_\text{I}$ magnetization possesses two values, alternating from shell to shell, which can be interpreted as single-site magnetizations of two emerging sublattices  [Fig.~\ref{mlow} (a)]. Although this behavior has been studied quite in detail for a simple Ising model on a Husimi lattice  \cite{comm, jetp, monroe}, it results in a novel feature that we find here. Namely, the Heisenberg $m_\text{H}$ magnetization bifurcates as well, that is due to the consistency of the both subsystems. In other words, despite the ground state of the Heisenberg trimer being rigid, its environment (i.e., the Ising-type spins that constitute the Husimi lattice) brings about drastic changes in its thermodynamic properties, resulting in a sublattice structure of the Heisenberg system, as shown in the inset of Fig.~\ref{mlow}(a) (note also that this doubling bifurcation appears in the area of the formation of the above mentioned plateau at $m_\text{H}=1/6$). More interestingly, as we continue decreasing the temperature, we find further doubling bifurcations both for the Heisenberg and for the Ising magnetizations  $-$ new bubbles are formed as parts of the old ones, resulting in $2^m$ periodic phases ($m=2, 3, ...$), with more complicated sublattice structure.

Finally, for ultimately low temperatures the system undergoes a transition to a chaotic phase, where the magnetization pattern does not repeat itself anymore. We emphasize again that the full period-doubling cascade and the chaotic behavior appears not only for the Ising magnetization, but also for the Heisenberg one, which will consequently have a crucial impact on the entanglement behavior as well. Moreover, the chaos ensues in the vicinity of the incipient magnetization plateau at $m_\text{H}=1/6$, i.e., where the above mentioned highly entangled $W$ state is the ground state of the Heisenberg trimer.

Furthermore, the chaotic regime contains $p$-periodic windows ($p=3, 5, ... $) with $p\cdot 2^m$ periodic phases. Particularly a wide three-periodic window is plainly distinguishable in Fig.~\ref{mlow}(b). Note also that the transition from and to chaos at both edges of the window takes place by means of a tangent bifurcation $-$ behavior that is quite different from that of usual polynomial mappings, such as the logistic one. This peculiar feature of maps describing spin-lattice models has been discussed in our previous works (note that the $Q$-state Potts model on a Bethe lattice exhibits similar behavior versus the temperature $T$ and for noninteger values of $Q<2$) \cite{comm, jetp}.

Meanwhile, to show explicitly the appearance of chaos, and to distinguish more rigorously between long-periodic phases and truly chaotic behavior, we consider the Lyapunov $\lambda(x)$ exponent. The latter tells one whether an infinitesimal perturbation in initial conditions has an infinitesimal effect ($\lambda(x)<0$ $-$ periodic behavior), or leads to a totally different trajectory ($\lambda(x)>0$), that would correspond to a chaotic regime (note that at a bifurcation point $\lambda(x)=0$). The Lyapunov exponent is defined as:
\begin{equation}\label{21}
\lambda{(x)}=\lim_{m\rightarrow\infty}{\frac{1}{m}}\ln\left|\frac{df^{(m)}(x)}{dx}\right|,
\end{equation}
where $f^{(m)}(x)$ stands for the $m$th iteration of the mapping $f(x)$.

\begin{figure}[ht]
\begin{center}
\includegraphics[width=7cm]{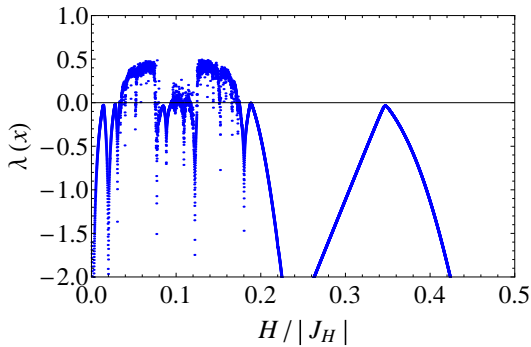}
\caption {(Color online) The Lyapunov exponent $\lambda(x)$ of the mapping $f(x)$ [Eq.~(\ref{16})], versus the magnetic field $H_\text{H}=H_\text{I}\equiv H$. Here $J_{\text{IH}}/J_{\text{H}}=-0.01$, $J_3/J_\text{H}=0.5$, $T/|J_\text{H}|=0.02$ and $q=3$ ($J_{\text{H}}^{xy}=J_{\text{H}}^{zz}\equiv J_{\text{H}}$) [the values of the parameters are the same as in Fig.~\ref{mlow}(b)]. \label{lyap}}
\end{center}
\end{figure}

As Fig.~\ref{lyap} shows, the system indeed exhibits chaos, that corresponds to the areas where $\lambda(x)>0$. Additionally, the above mentioned tangent bifurcations, corresponding to a transition from a chaotic phase to a periodic window, are also seen here [the periodic windows are the inclusions of negative Lyapunov exponent inside the chaotic phase; cf. Fig.~\ref{mlow}(b)]. Finally, the model possesses a variety of superstable cycles with $\lambda(x)=-\infty$ \cite{lee1}: as was shown in Ref.~\cite{comm}, at these points a specific change occurs in the symbolic dynamics of the mapping (\ref{16}).

\subsection{Entanglement}\label{entanglb}

Before starting the analysis of the entanglement properties of the Heisenberg trimer of the THL, we note that a system of three qubits has been already studied from different perspectives and in various environments, including, e.g., external \cite{solomon}, or effective magnetic fields (that may describe a specific lattice structure \cite{me22}), incident \cite{me1} or cavity light field dressing \cite{me2}, etc., and therefore is quite well understood by now. Nevertheless, the recursive structure, bringing about chaotic behavior that we present below, has not been considered yet. Therefore, we do not stop much in detail on the analysis of the ground-state features of the trimer of the THL, but proceed, instead, to the studies of its periodic  and chaotic entangled regimes.

\begin{figure}[ht]
\begin{center}
\small(a) \includegraphics[width=7cm]{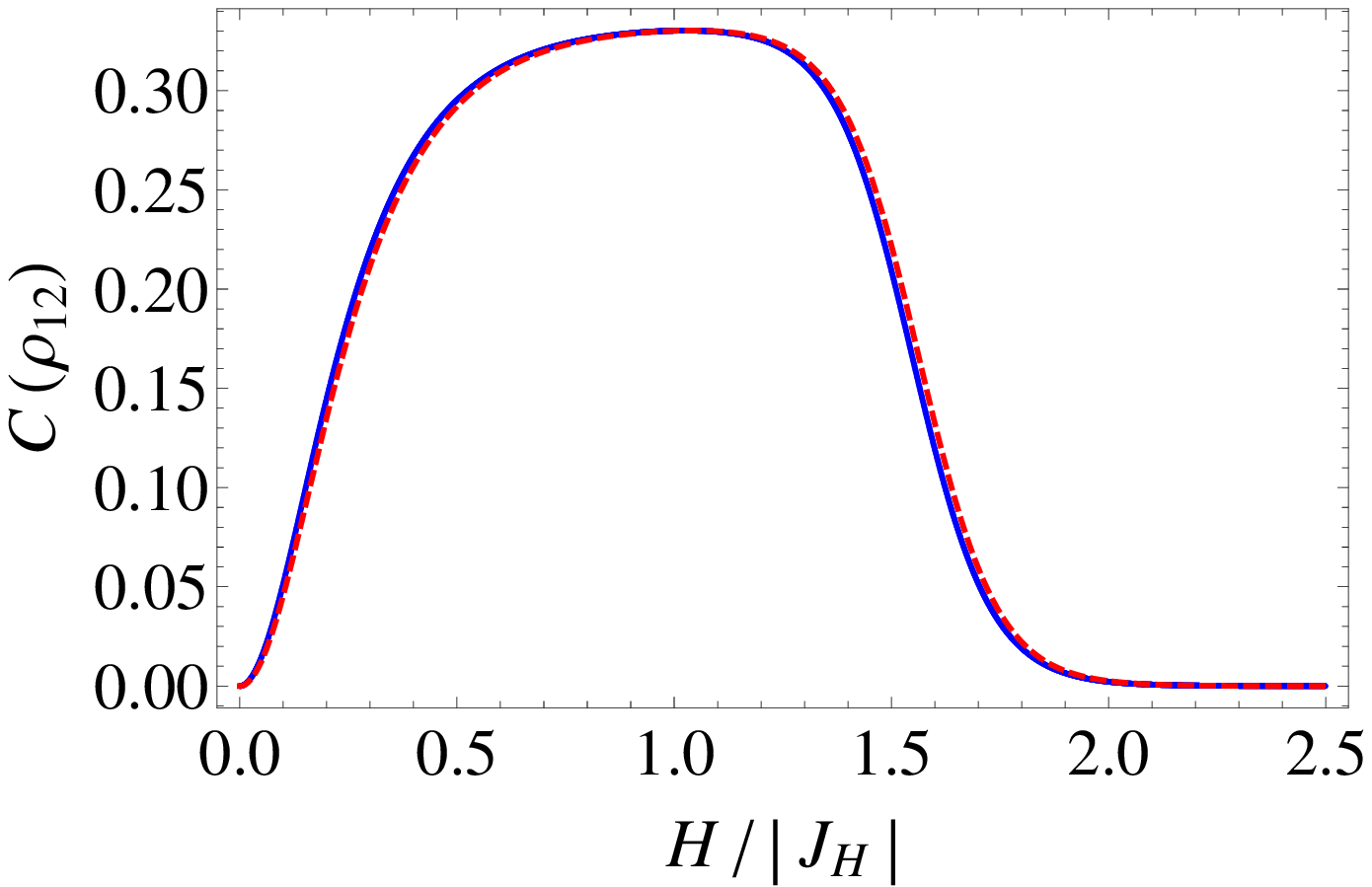}\\
\small(b) \includegraphics[width=7cm]{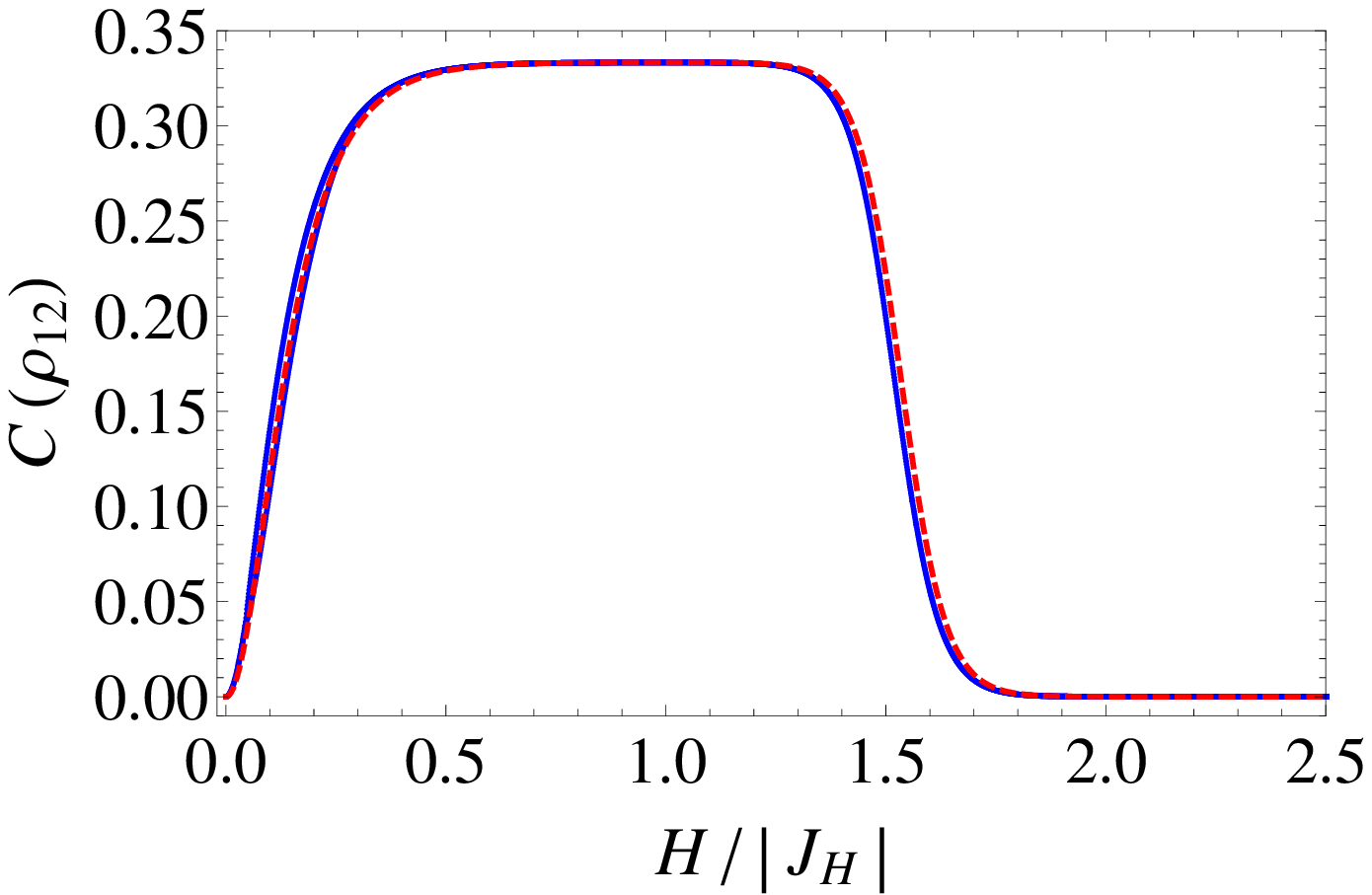}\\
\small(c) \includegraphics[width=7cm]{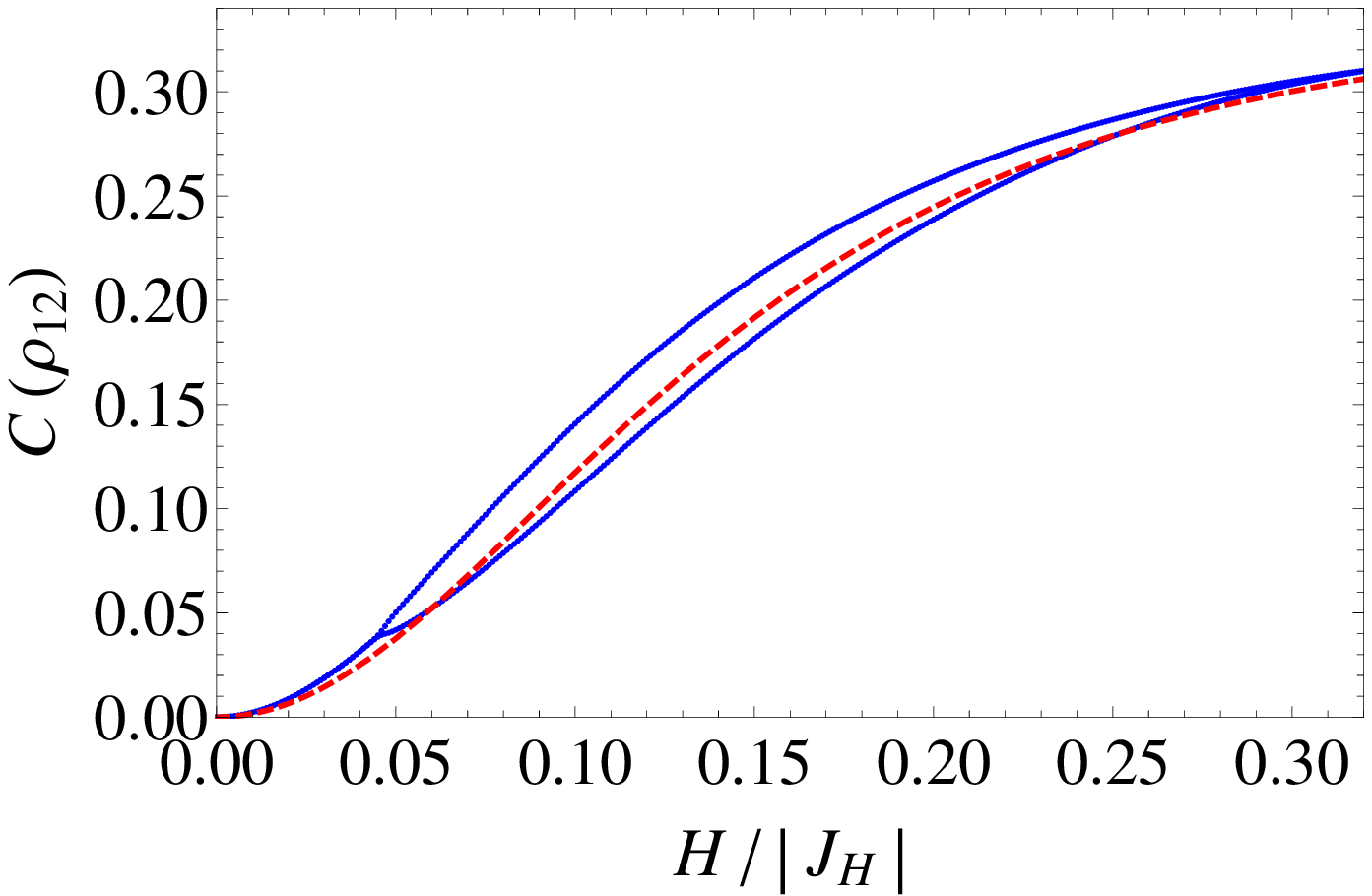}\\
\caption {(Color online) The concurrence of a pair of qubits from a Heisenberg trimer versus the magnetic field. The full blue curves correspond to the concurrence $C(\rho_{12})$ of our model, given by Eq.~(\ref{20}) ($H_\text{H}=H_\text{I}\equiv H$, $J_{\text{H}}^{xy}=J_{\text{H}}^{zz}\equiv J_{\text{H}}$), while the dashed red curves stand for the concurrence of a simple Heisenberg trimer with two-site isotropic interactions of a strength $J_\text{H}$, in an external magnetic field $H$ (cf.~\cite{solomon}). Here $J_{\text{IH}}/J_{\text{H}}=-0.01$, $J_3/J_\text{H}=0.5$, $q=3$ and (a) $T/|J_\text{H}|=0.09$; (b) $T/|J_\text{H}|=0.05$; (c) Enlargement of the period doubling zone shown in (b). \label{conc}}
\end{center}
\end{figure}

Having the exact formula for the concurrence $C(\rho_{12})$ of a pair of qubits from the Heisenberg trimer, expressed in terms of the magnetization and spin-spin correlation functions [Eq.~(\ref{20})], we plot it in Fig.~\ref{conc}, as a function of the magnetic field $H_\text{H}=H_\text{I}\equiv H$. As Fig.~\ref{conc}(a) shows, the overall behavior of the concurrence at relatively high temperatures and weak Ising-to-Heisenberg coupling is quite analogous to that of a simple system of just three qubits in an external magnetic field: the figure compares the pairwise entanglement of a pure Heisenberg trimer (dashed red curve), with that of our model (full blue curve). This similarity becomes evident from the fact that the coupling to the rest of the system, and particularly, to the Ising spins that constitute the Husimi lattice, happens by means of the weak $J_\text{IH}$ interaction. Meanwhile, as for the broad maximum at around $C(\rho_{12})=1/3$, it appears due to the ground $W$ state $|\psi\mathrm\rangle=\frac{1}{\sqrt{3}}\left(|\uparrow\uparrow\downarrow\rangle+|\uparrow\downarrow\uparrow\rangle+|\downarrow\uparrow\uparrow\rangle\right)$, that passes to a saturated non-entangled state $|\uparrow\uparrow\uparrow\rangle$ (the thermal effects smooth here the abrupt jump from $C(\rho_{12})=1/3$ to $C(\rho_{12})=0$). Note, however, that as $J_\text{IH}/J_\text{H}$ ratio is increased, the transition from the $W$ to the saturated state is shifted to smaller values of $H$, since the Ising-to-Heisenberg coupling can be interpreted as an additional magnetic field of a strength $J_\text{IH}\mu_k$ ($\mu_k=\pm3/2, \pm1/2$), acting upon the qubits [see also the Hamiltonian (\ref{1})].

Nevertheless, the features of our model at sufficiently low temperatures are strikingly different from the above picture. Namely, the entanglement of the Heisenberg trimer bifurcates as the temperature is decreased, even when the Ising-to-Heisenberg coupling is weak [see Fig.~\ref{conc}(b), and its details in the area of period doubling in Fig.~\ref{conc}(c)]. This means that the concurrence $C(\rho_{12})$ has now two values, that interchange one with another from shell to shell. In other words, we find here a rise of two sublattices with different values of {\it pairwise entanglement}. Moreover, the breaking up into sublattices with respect to the entanglement is identical to that with respect to the magnetization $m_\text{H}$: the upper branch of $m_\text{H}$ in Fig.~\ref{mlow}(a) corresponds to the upper branch of $C(\rho_{12})$ in Fig.~\ref{conc}(c). It is also worth noting that the bifurcation occurs in the weak magnetic field region, where the {\it magnetic entanglement} is induced (the ground state at $H=0$ is not entangled, and quantum correlations build up here due to the inclusion of the magnetic field \cite{bose}).

\begin{figure}[ht]
\begin{center}
\small(a) \includegraphics[width=7cm]{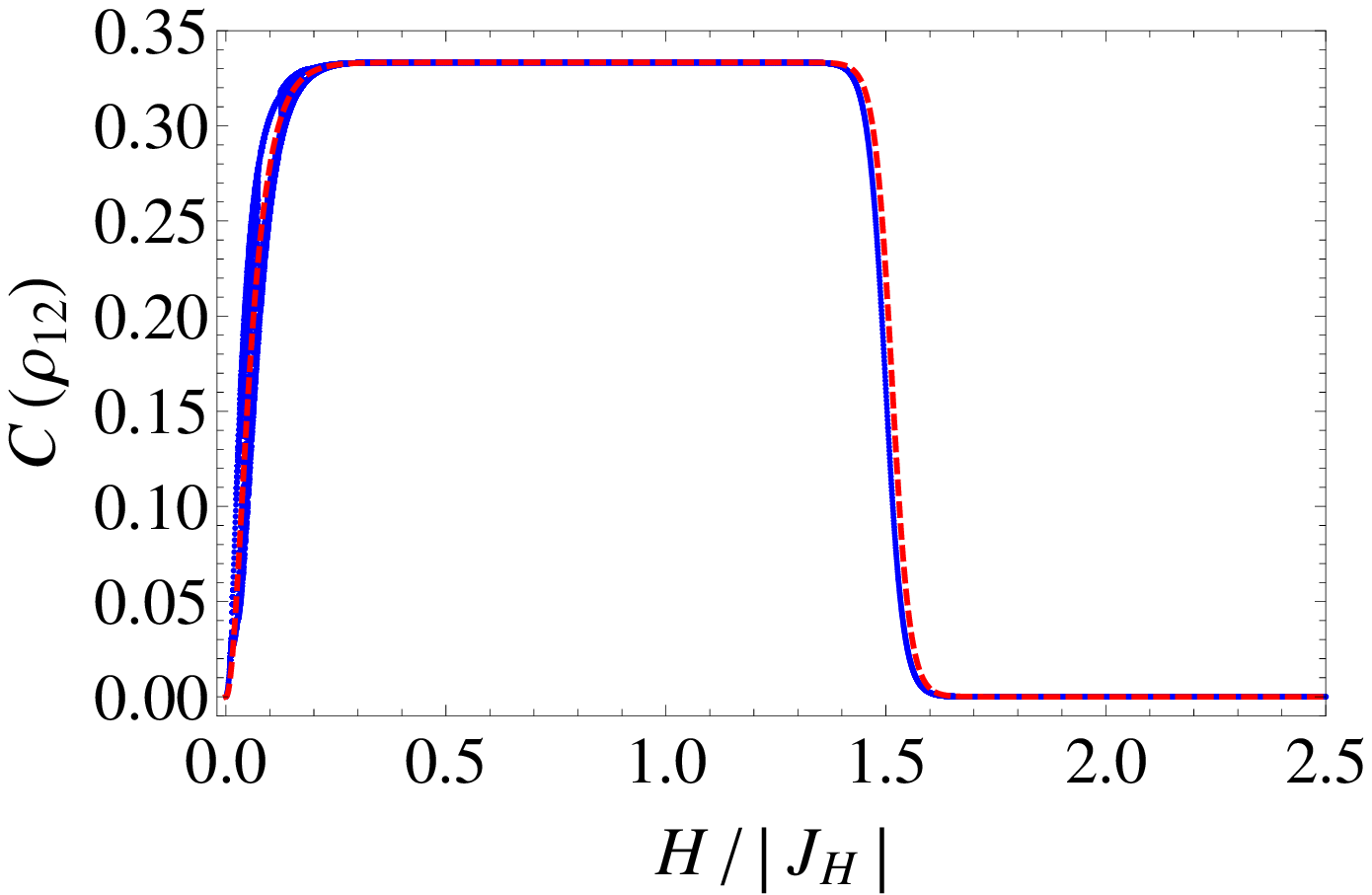}\\
\small(b) \includegraphics[width=7cm]{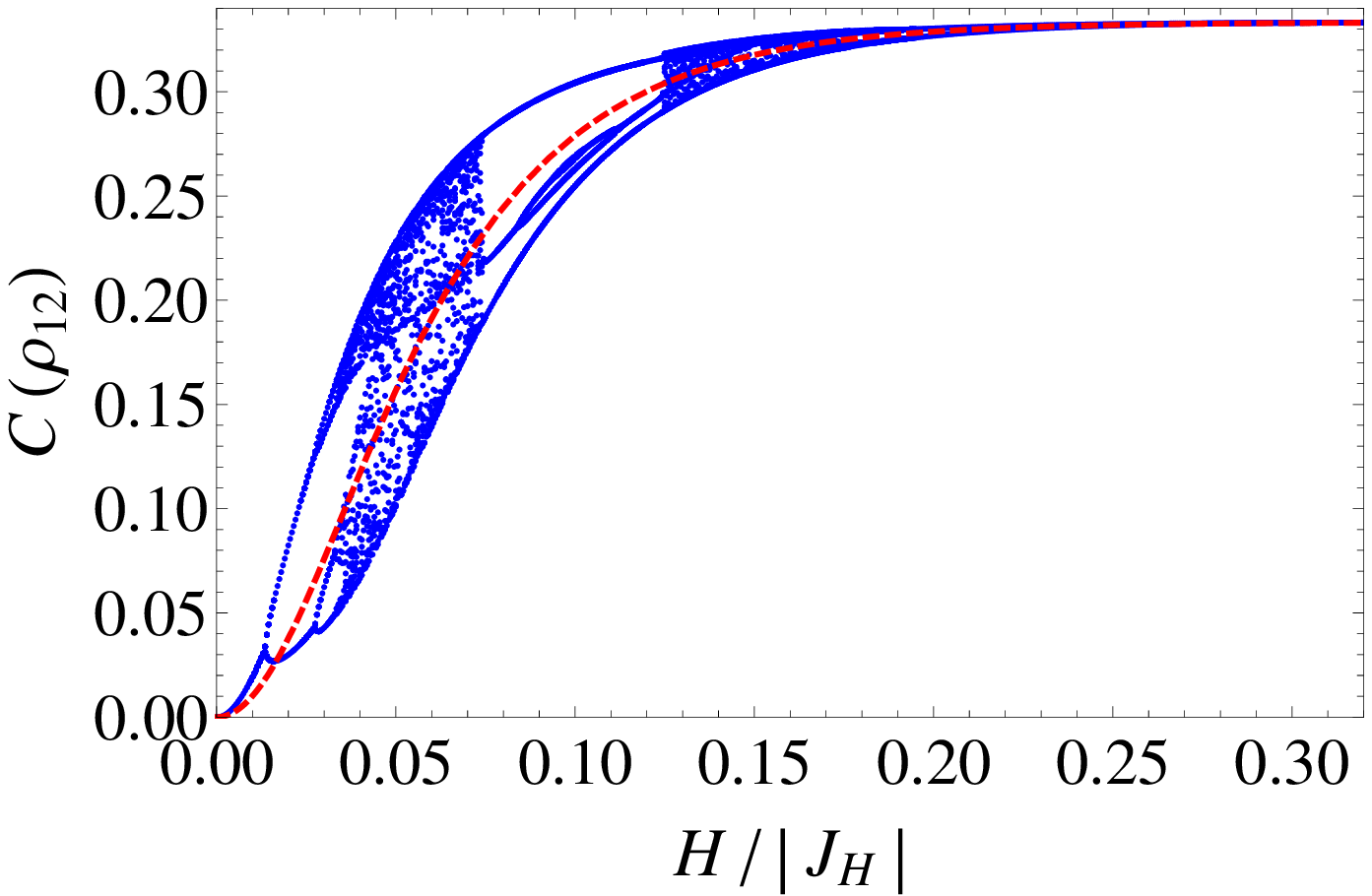}\\
\caption {(Color online) (a) The concurrence of a pair of qubits from a Heisenberg trimer versus the magnetic field. The full blue curve corresponds to the concurrence $C(\rho_{12})$
of our model, given by Eq.~(\ref{20}) ($H_\text{H}=H_\text{I}\equiv H$, $J_{\text{H}}^{xy}=J_{\text{H}}^{zz}\equiv J_{\text{H}}$), while the dashed red curve stands for the concurrence
of a simple Heisenberg trimer with two-site isotropic interactions of a strength $J_\text{H}$, in an external magnetic field $H$. Here $J_{\text{IH}}/J_{\text{H}}=-0.01$, $J_3/J_\text{H}=0.5$,
$T/|J_\text{H}|=0.02$ and $q=3$; (b) Enlargement of the chaotic zone shown in (a). \label{conchaos}}
\end{center}
\end{figure}

Meanwhile, at ultimately low temperatures, in the same manner as for the magnetizations $m_\text{I}$ and $m_\text{H}$, the concurrence exhibits more and more phases with higher $2^m$-periods  ($m=2, 3, ...$), and eventually, one reaches the regime of {\it chaotic entanglement} (Fig.~\ref{conchaos}). This means that in the entire system (deep inside the Husimi tree) the sequence of the values of the concurrence does not repeat itself, and is, in other words, incommensurate. We also note that the above chaotic behavior appears solely due to the specific $-$ recursive $-$ structure of the THL, and due to the presence of three-site interactions in the initial (\ref{1}), and therefore, in the effective Ising  (\ref{8}) model (as is known, the Ising model on a Husimi lattice with only two-site interactions does not exhibit chaos \cite{monroe, monroe1}). Moreover, the region of the chaotic entanglement is directly related to the strength of the three-site interactions $-$ the stronger are these interactions, the wider is the chaotic region.

On the other hand, we note that purely thermal entanglement, i.e., entanglement that arises at $H=0$, only due to the thermal mixing of the eigenstates of the system, exhibits neither period doubling, nor chaotic behavior. In other words, chaos ensues only in the presence of the external magnetic field, where an entangled state appears in the ground-state structure. Furthermore, chaos does not induce entanglement by itself, but may enhance (or reduce) its amount from shell to shell. Specifically, as shown in Fig. ~\ref{conchaos}(b), the chaotic entanglement is dominantly stronger than that of just three qubits in an external magnetic field (compare the full blue and dashed red curves therein).

Furthermore, an important observation is that the Lyapunov $\lambda(x)$ exponent [Eq.~(\ref{21})], that describes the system's chaotic behavior, characterizes both the magnetic and the entanglement quantities (e.g., the magnetization and the concurrence) of the present model. More precisely, as already mentioned in Sec.~\ref{recursive}, the properties of the system in the thermodynamic limit are defined by the mapping (\ref{16}), as particularly are the per-site magnetizations $m_\text{H}$ and $m_\text{I}$, as well as the pairwise entanglement of the Heisenberg trimer, expressed in terms of $C(\rho_{12})$. Therefore the $\lambda(x)$ exponent of the mapping (\ref{16}), plotted in Fig.~\ref{lyap}, confirms and quantifies the above discussed chaotic behavior of the spin-spin entanglement, shown in Fig.~\ref{conchaos}. Moreover, the inclusion of $\lambda(x)<0$ regions inside the chaotic regime, and $\lambda(x)=-\infty$ points in Fig.~\ref{lyap} reveal the existence of periodic windows and superstable cycles of the {\it entanglement} in the present model (similar to that of the magnetization, as discussed in the previous subsection).

\begin{figure}[ht]
\begin{center}
\includegraphics[width=7cm]{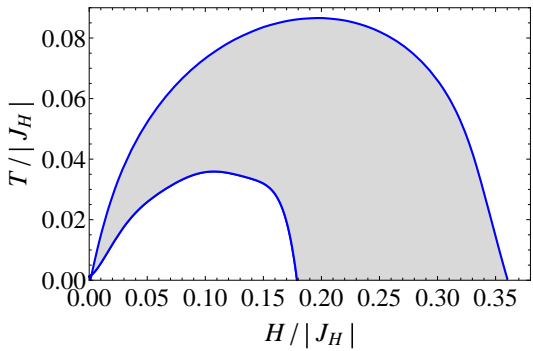}
\caption {(Color online) The phase transition lines from the uniform phase to a two-periodic one (the upper curve) and from periodic to chaotic regimes (the lower curve). The full period-doubling cascade takes place in the shaded area between these two curves. Below the lower curve the chaotic entanglement ensues, which may be interrupted with periodic windows. Here $J_{\text{IH}}/J_{\text{H}}=-0.01$, $J_3/J_\text{H}=0.5$, and $q=3$ ($J_{\text{H}}^{xy}=J_{\text{H}}^{zz}\equiv J_{\text{H}}$). \label{phase}}
\end{center}
\end{figure}

Finally, for more details, in Fig.~\ref{phase} we show the transition line between the phase without sublattice structure (uniform phase) and the 2-periodic one on the one hand,
and the transition line between periodic and chaotic regimes on the other hand. More precisely, the upper line in Fig.~\ref{phase} corresponds to the first bifurcation point,
and up to the lower line (the shaded region) one has the full $2^m$ ($m=1, 2, 3, ...$) period doubling cascade with periodic entanglement behavior. Meanwhile below the lower line the system exhibits aperiodic chaotic entanglement, that, however, may be interrupted with $p$-periodic windows ($p=3, 5, ...$).

\section{Conclusion}\label{concl}

We proposed a multisite interaction spin-1/2 Ising-Heisenberg model on a triangulated Husimi lattice to study the effects of chaos and bifurcation on the system's entanglement features. We used the generalized star-triangle transformation to map the initial model onto an effective Ising one on a simple Husimi lattice, which we then solved by means of the recursive method. An exact formula was obtained for the concurrence, to quantify the pairwise spin-spin entanglement of the embedded Heisenberg triangles, expressed through the system's single-site magnetization and spin-spin correlation functions.

We have shown that at relatively high temperatures and weak Ising-to-Heisenberg coupling, the model exhibits quite usual magnetic behavior and its entanglement properties are analogous to that of just three intercoupled qubits. However, one finds drastic changes in the properties of the system as the temperature $T$ is decreased. Namely, period doubling of per-site Ising and Heisenberg magnetizations appears such that the corresponding values interchange one with another from shell to shell (this corresponds to a bifurcation point of the mapping that describes our model). We interpret this behavior as a rise of a sublattice structure, that affects the system's entanglement properties, too: the concurrence now exhibits period doubling as well, and each of the two values of the entanglement is related to a particular magnetic sublattice. In other words, here every branch of the concurrence corresponds to a specific branch of the Ising and Heisenberg single-site magnetizations.

On the other hand, with the further decrease in the temperature, we find a full-period doubling cascade, and eventually a transition to a chaotic phase where the sequence of the entanglement (and magnetization) values is aperiodic and does not repeat itself. Importantly, this behavior appears due to the inclusion of three-site interactions $-$ the system is always periodic when only pair interactions are taken into account. It is also interesting to note that chaos ensues in the region of incipient magnetic and concurrence plateaus, that correspond to the entangled $W$ ground state. Additionally, chaotic behavior is absent if the system exhibits only thermal entanglement (in the absence of the magnetic field), i.e., if its ground state is separable, which, therefore, along with the above argument, points to a connection of ground-state entanglement and chaos. Moreover, although chaos does not induce entanglement by itself, it however may slightly enhance the latter (somewhat similar results, namely entanglement enhancement for chaotic initial conditions, have been also indicated, e.g., in Ref.~\cite{dicke}, although within distinct mechanisms for the chaos development). It is important to note, however, that both for $2^m$-periodic ($m=1, 2, ...$) and for the chaotic regimes the ground state of the Heisenberg trimer is rigid, whereas the Ising sublattice, that is (weakly) coupled to the Heisenberg triangles, induces the underlying complex behavior. Consequently, at sufficiently low temperatures the entanglement (as well as the magnetization) of the Heisenberg spins exhibits a rich phase structure, including the above mentioned $2^m$-periodic phases, chaos and $p$-periodic windows ($p=3, 5, ...$), as well as various superstable cycles, that we have also confirmed by the studies of the Lyapunov $\lambda(x)$ exponent. The above transitions between periodic phases occur through doubling bifurcations, whereas back and forth transitions from chaos to periodic windows happen through tangent bifurcations, that we plan to study in more detail in our future works.

Finally, we emphasize that for the explicit studies of the chaotic behavior of the above spin model and specifically, the chaotic pairwise entanglement of its Heisenberg trimer, we have mainly used the Lyapunov exponent approach. The latter allows one to differentiate rigorously between periodic and chaotic regimes and to measure the strength of the emerging chaos. In this respect, our choice of the Lyapunov exponent characteristics has been particularly reinforced by the fact that the properties of the present spin model (and its state in the thermodynamic limit) are defined by a one-dimensional recursive mapping. As is known, in such a case the Lyapunov $\lambda(x)$ exponent serves as a strong tool for the analysis of the inherent chaotic (as well as periodic) regimes \cite{comm, jetp, lyap}, and moreover, for the model that we studied here, it characterizes the chaotic behavior not only of the system's magnetic, but also of its entanglement features. Nevertheless, it is worth noting that for a deeper understanding of the relation between entanglement and chaos, apart from the Lyapunov exponent studies, one would necessitate here additional analysis, as, e.g., the estimation of the ergodicity of wave functions and level spacing statistics, quantum purity of the emerging mixed $X$ states, etc. (see, e.g. Refs.~\cite{dima1, dima2, kruch}): an important issue that we will address in our future work.

\section*{acknowledgments}

The authors are grateful to Thierry Platini, Nerses Ananikian and Nikolay Izmailyan for useful discussions. This research was conducted within the scope of the International Associated Laboratory (CNRS-France \& SCS-Armenia) IRMAS. We acknowledge additional support from GA-INCO-295025-IPERA FP7 program. L.C. gratefully acknowledges the support from the Conseil R\'{e}gional de Bourgogne.

\end{document}